\documentclass[letterpaper,aps,prl,twocolumn,showpacs]{revtex4}
\usepackage{graphicx,bm}
\usepackage{amsmath,amssymb}

\newcommand{\mysw}[1]{{\scriptscriptstyle #1}}

\begin{document}

\title{Renormalization-group exponents for superconducting phases in two-leg ladders}
\author{Yiwei Cai$^1$, Wen-Min Huang$^{2}$ and Hsiu-Hau Lin$^{1}$}
\affiliation{$^1$Department of Physics, National Tsing Hua University, Hsinchu 30013, Taiwan\\
$^2$Physics Division, National Center for Theoretical Sciences, Hsinchu 30013, Taiwan}
\date{\today}

\begin{abstract}
In previous studies, we proposed a scaling ansatz for electron-electron interactions under renormalization group transformation. With the inclusion of phonon-mediated interactions, we show that the scaling ansatz, characterized by the divergent logarithmic length $l_d$ and a set of renormalization-group exponents, also works rather well. The superconducting phases in a doped two-leg ladder are studied and classified by these renormalization-group exponents as demonstration. Finally, non-trivial constraints among the exponents are derived and explained.
\end{abstract}


%71.10.Fd	Lattice fermion models (Hubbard model, etc.)

%71.10.Hf Non-Fermi-liquid ground states, electron phase diagrams and phase transitions in model systems

%71.27.+a	Strongly correlated electron systems; heavy fermions

%71.10.Pm Fermions in reduced dimensions (anyons, composite fermions, Luttinger liquid, etc.) (for anyon mechanism in superconductors, see 74.20.Mn)

%74.20.Mn Nonconventional mechanisms (spin fluctuations, polarons and bipolarons, resonating valence bond model, anyon mechanism, marginal Fermi liquid, Luttinger liquid, etc.)

%74.20.Mn	 Nonconventional mechanisms (spin fluctuations, polarons and bipolarons, resonating valence bond model, anyon mechanism, marginal Fermi liquid, Luttinger liquid, etc.)

\maketitle
Even though superconductivity has been observed\cite{Onnes1911} for one hundred years, the mechanism turning electrons into pairs remains an open question. In conventional superconductors, phonons mediate effective attractions\cite{BCS1957} between electron in the low-energy limit and lead to Cooper pair formation near the Fermi surface. On the other hand, for unconventional superconductors\cite{Norman2011} like cuprates, phonon-mediated interactions seem to play a secondary role while the electronic correlations are believed to reigns. It is speculated that the spin fluctuations is the key to the pairing mechanism (and perhaps the pairing symmetry as well). However, it hasn't been fully understood how Coulomb repulsion eventually glues electrons into pairs. 

The discovery of iron-based superconductors keeps the puzzling charm going. Collecting from experimental observations, it is believed that electronic correlations in these materials are important but much weaker than those in cuprates. Though the renormalization-group (RG) analysis\cite{Chubukov08,Wang2009,Wang2011} for electron-electron interactions delivers the correct pairing symmetry, the isotope effects measured in laboratories show conflicting results. In addition, the iron-based superconductors are distinct from the cuprates due to the presence of multiple active bands. 

Inspired by the anomalous isotope effects in iron-based superconductors, it is helpful to investigate the competitions between electron-electron\cite{Shankar1994,Honerkamp2001} and electron-phonon interactions\cite{Zimanyi1988,Bindloss2005,Seidel2005} by RG analysis. The major difficulty lies in the enormous couplings for all allowed interactions. Besides, as the number of couplings grow, reading out the desired messages from RG flows can be challenging as well. Recently, we found a scaling ansatz\cite{Shih2010}, characterized by a set of RG exponents, for electron-electron interactions in many correlated systems. These RG exponents build a clear hierarchy of relevant couplings and serve as an unambiguous indicator for the ground-state instabilities. We are curious whether similar scaling ansatz exists even when the retarded interactions mediated by phonons are included.

\begin{figure}
\centering
\includegraphics[width=\columnwidth]{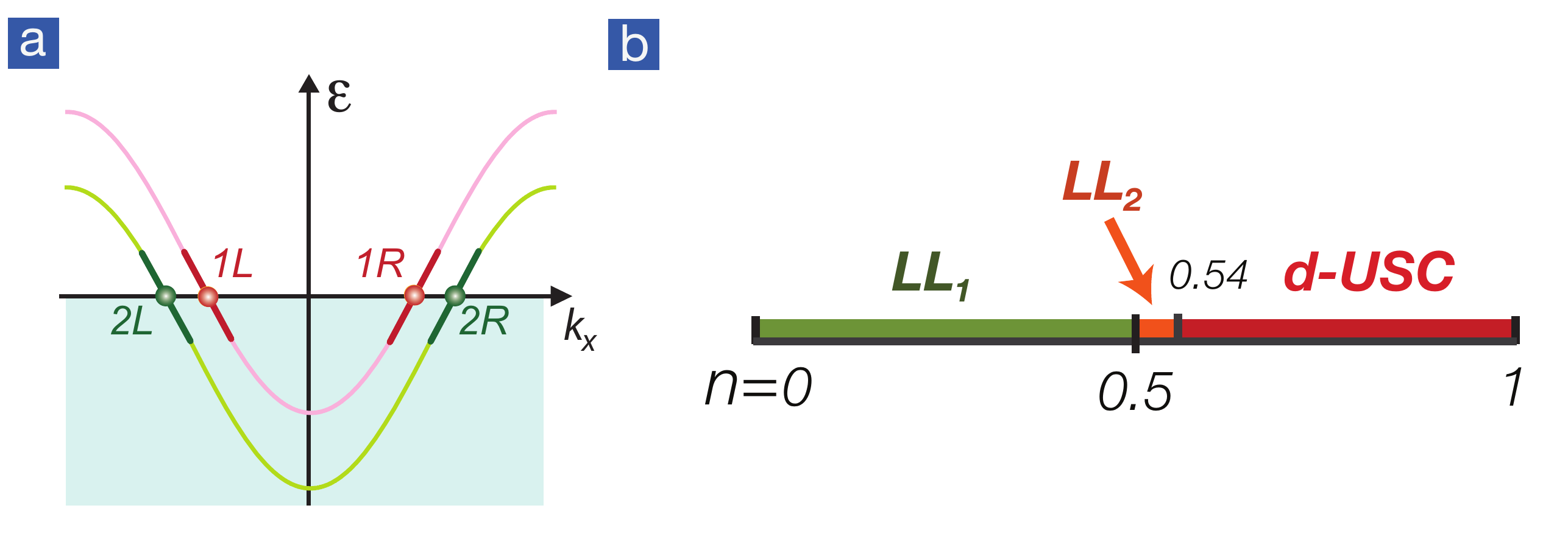}
\caption{(a) Band structure for a two-leg ladder at filling $0<n<1$. The low-energy excitations are described by four pairs of chiral fields $\psi_{Ri\alpha}, \psi_{Li\alpha}$ with $i=1,2$ and $\alpha=\uparrow,\downarrow$. (b) Phase diagram for a doped two-leg ladder. In the filling regime $0.5<n\lesssim 0.54$, the interband Cooper scattering becomes irrelevant and the superconducting phase decomposes into the 2-channel Luttinger liquid.}
\end{figure}

In this Rapid Communications, we show that the scaling ansatz works in the presence of both electron-electron and electron-phonon interactions and the patterns of the extracted RG exponents describe the ground states and also the quantum phase transitions between them. To make the discussions concrete, we study the exemplar two-leg ladder shown in Fig. 1. The low-energy excitations are described by four pairs of chiral fields $\psi_{\mysw{R}i\alpha}, \psi_{\mysw{L}i\alpha}$, where $R/L$ denotes the chirality and $i=1,2$, $\alpha=\uparrow,\downarrow$ are the band and spin indices. The generic interactions can be classified into two categories: forward and Cooper, described by the following effective action\cite{Fabrizio1993,Balents1996,Schulz1996,Arrigoni1996,Lin1997,Lin1998,Chang2005,Szirmai2006,Lin2008},
\begin{eqnarray}
S_{\rm int} &=& \int d\tau \int dx \sum_{i,j=1,2}
\sum_{\alpha,\beta=\uparrow,\downarrow} \pi (v_i+v_j) \times
\nonumber\\
&& \hspace{-15mm} \bigg[
-f^{\ell}_{ij} \psi^\dag_{\mysw{R}i\alpha} \psi^{}_{\mysw{R}i\beta}
\psi^\dagger_{\mysw{L}j\beta}\psi^{}_{\mysw{L}j\alpha}
+f^{s}_{ij} \psi^\dag_{\mysw{R}i\alpha} \psi^{}_{\mysw{R}i\alpha}
\psi^\dagger_{\mysw{L}j\beta}\psi^{}_{\mysw{L}j\beta}
\nonumber\\
&& \hspace{-12mm}
- c^{\ell}_{ij} \psi^\dag_{\mysw{R}i\alpha} \psi^\dag_{\mysw{L}i\beta}
\psi^{}_{\mysw{L}j\alpha} \psi^{}_{\mysw{R}j\beta}
+c^{s}_{ij} \psi^\dag_{\mysw{R}i\alpha} \psi^\dag_{\mysw{L}i\beta}
\psi^{}_{\mysw{L}j\beta} \psi^{}_{\mysw{R}j\alpha} \bigg],
\label{H}
\end{eqnarray}
where $f_{ij}$ and $c_{ij}$ denote the forward and Cooper scattering between the $i-$th and $j-$th bands. To avoid double counting, the diagonal parts of the forward scattering are set to zero, $f_{ii} \equiv 0$. The superscripts $\ell, s$ stand for large and small momentum transfer between chiral fields of the same spin index.

The ground state of the doped two-leg ladder\cite{Balents1996,Krotov1997,Shih2010} depends on the filling $n$, i.e. average number of electrons per lattice site. For $0.54 \lesssim n<1$, electrons form pairs and the ground state is an unconventional superconductor (with power-law correlations). The wave functions in different bands reveal opposite signs, resembling the pairing symmetry in iron-based superconductors. That is to say, the pairing wave functions at $k_y=0$ (band 2) and $k_y=\pi$ (band 1) are opposite in signs and thus referred as ``d-wave" unconventional superconductor (d-USC). In the filling interval, $0.5<n\lesssim 0.54$, the pairing instability disappears and the ground state is a 2-channel Luttinger liquid. Bellow the filling $n<0.5$, only one band remains active and the low-energy physics is well described by the single-channel Luttinger liquid. Despite the simplicity of the phase diagram, it is worth mentioning that RG exponents for the relevant couplings help to clear up some confusions in the literature.

Electron-phonon interactions are different in nature and retardation effects must be taken care of. To include both electron-electron and electron-phonon interactions, it requires turing the couplings $g_i$ into frequency-dependent functions $g_i(\omega)$, making the RG analysis challengingly difficult. The old trick in the original Bardeen-Cooper-Schrieffer (BCS) theory comes to rescue -- we assume the phonon-mediated interactions only occur within a thin shell, roughly the order of Debye frequency $\omega_D$, near the Fermi surface. Within this approximation, the frequency-dependent couplings take the following form,
\begin{align}
g_i(\omega) = g_i + \Theta(\omega_D-\omega) \tilde{g_i},
\end{align}
where $g_i$ and $\tilde{g}_i$ are the (instantaneous) electron-electron and the (retarded) phonon-mediated interactions. The step-like shape is invariant under RG transformation. Let $1$ and $\Theta$ represent the shapes of constant and step respectively. It is clear that the algebra of the shapes is closed: $1 \times 1 =1$, $1 \times \Theta = \Theta \times 1 = \Theta$ and $\Theta \times \Theta = \Theta$. As a consequence, the renormalized couplings maintain the same step-like shapes and make the approximate frequency dependence self-consistent.

\begin{figure}
\centering
\includegraphics[width=7cm]{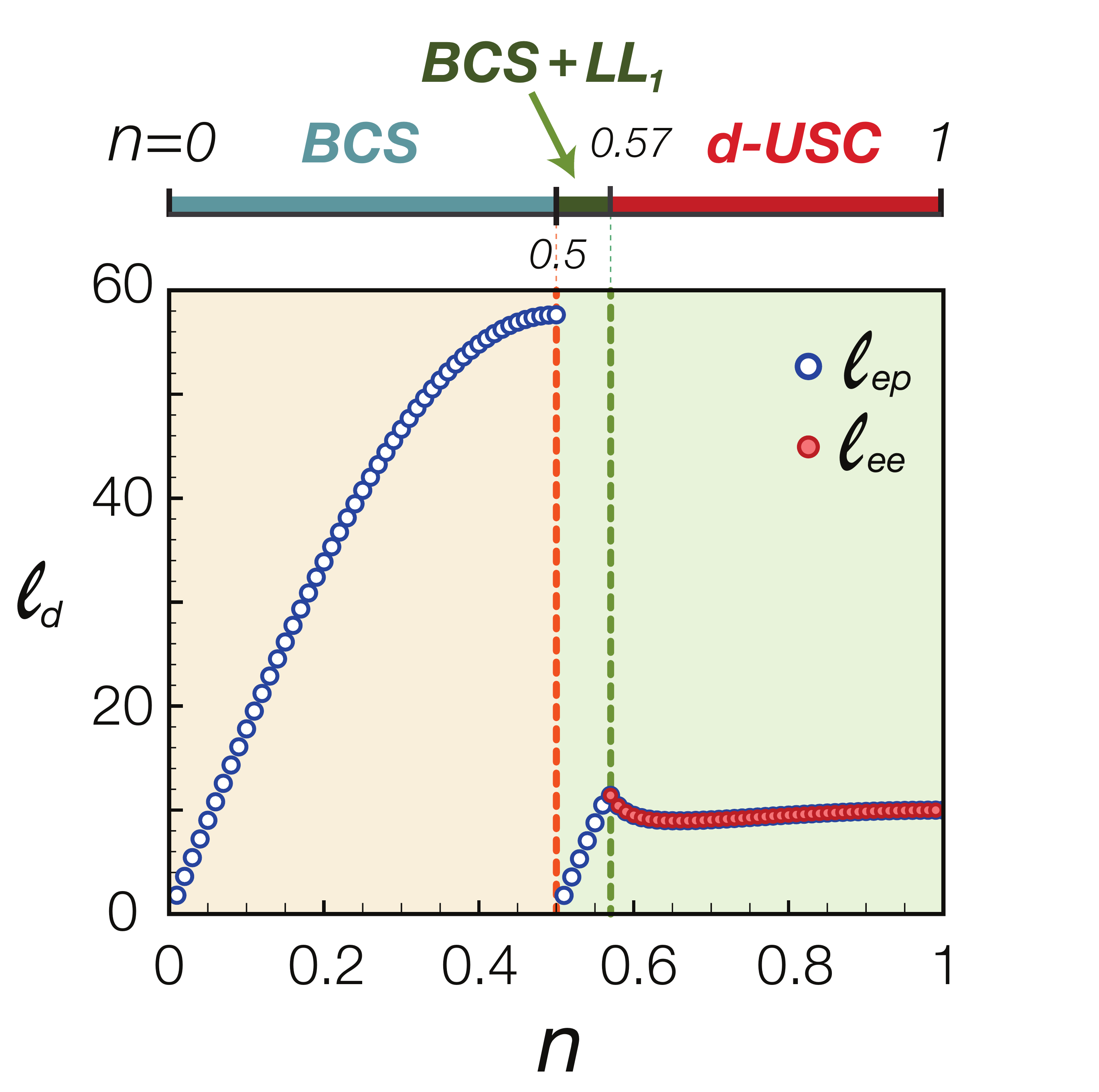}
\caption{Phase diagram for a doped two-leg ladder with both electron-electron and electron-phonon interactions. The critical filling $n_c \approx 0.57$ separates two types of superconducting instabilities driven by electron-electron interactions and phonon-mediated ones.}
\end{figure}

Following standard derivations, the RG equations for the retarded interactions are
\begin{align}
\partial_l \tilde{c}^\ell_{ii} =& 
-2 (\tilde{c}^\ell_{ii})^2 
-4 c^\ell_{ii} \tilde{c}^\ell_{ii}
+2 c^s_{ii} \tilde{c}^\ell_{ii},
\nonumber\\
\partial_l \tilde{c}^\ell_{ij} =&
-4 \tilde{f}^\ell_{ij} \tilde{c}^\ell_{ij}
-4 c^\ell_{ij} \tilde{f}^\ell_{ij}
+2 c^s_{ij} \tilde{f}^\ell_{ij}
-4 f^\ell_{ij} \tilde{c}^\ell_{ij}
+2 f^s_{ij} \tilde{c}^\ell_{ij},
\nonumber\\
\partial_l \tilde{f}^\ell_{ij} =&
-2 (\tilde{f}^\ell_{ij})^2
-2 (\tilde{c}^\ell_{ij})^2
-4 f^\ell_{ij} \tilde{f}^\ell_{ij}
+2 f^s_{ij} \tilde{f}^\ell_{ij}
\nonumber\\
&-4 c^\ell_{ij} \tilde{c}^\ell_{ij}
+2 c^s_{ij} \tilde{c}^\ell_{ij}.
\label{rg-eqs}
\end{align}
The other set of retarded interactions $\tilde{c}^s_{ij}, \tilde{f}^s_{ij}$ does not renormalized at all within the approximation. The renormalized couplings $g_i(l)$ and $\tilde{g}_i(l)$ are solved numerically. It was found before that $g_i(l)$ is well captured by a scaling ansatz. It is rather remarkable that the numerical solutions for $\tilde{g}_i$ also follow a similar scaling ansatz,
\begin{eqnarray}
\tilde{g_i} \approx \frac{\tilde{G}_i}{(l_d-l)^{\gamma_{\tilde{g}_i}}},
\end{eqnarray}
where $l_d$ is the divergent (logarithmic) length, $\gamma_{\tilde{g}_i}$ is the RG exponent for the retarded coupling $\tilde{g}_i$ and $\tilde{G}_i$ is some non-universal order-one constant. We shall elaborate the importance of the divergent length $l_d$ and the RG exponents in later paragraphs. It is worth emphasizing that all renormalized couplings remain in the perturbative regime even though the ansatz mysteriously contains the divergent length scale $l_d$.

For numerical analysis, we choose the ratio $\tilde{g}_i/g_i = -0.1$, where the minus sign indicates the attractive nature of the phonon-mediated interactions. As shown in Fig. 2, an interesting critical filling $n_c \approx 0.57$ separates two types of superconductors driven by different pairing mechanisms. In the regime $n_c<n<1$, the instantaneous couplings $g_i(l)$ grow faster than the retarded ones $\tilde{g}_i(l)$ and the divergent length $l_d$ is dictated by the bare values of the electron-electron interactions. Compared with the phase diagram in Fig. 1, the superconducting ground state is qualitatively the same with unconventional pairing symmetry between different bands. In the regime $0<n<n_c$, the RG flows are different. Despite of the smaller bare values, the retarded couplings $\tilde{g}_i(l)$ reign and flow toward strong coupling first. The pairing mechanism is the well-known BCS theory and the isotope effect is expected to be significant. This regime is further divided into two parts because there is only one active band at lower fillings. In the two-band regime, $0.5<n<n_c$, Van Hove singularity develops in the almost empty band and the phonon-mediated Cooper scattering ($\tilde{c}_{11}$) drives the ground state superconducting. Compare with the phase diagram in Fig. 1, inclusion of the electron-phonon interactions causes the instability in the 2-channel Luttinger liquid and only one channel (in the majority band) survives. Further reduction in electron filling to the regime $n<0.5$, only the bottom band remains active. This is where our common intuitions work -- the repulsive electron-electron interactions (though with larger bare values) renormalize to zero while the attractive interactions mediated by phonons reign, rendering the ground state to the conventional BCS superconductor.

The critical density $n_c$ not only separates two different pairing mechanisms, but also marks different trends of the divergent length $l_d$ at different fillings. In the regime $0<n<n_c$, $l_d$ remains more or less constant when the filling varies. The constancy partially comes from the smooth evolution of Fermi velocities in this regime and is partially attributed to the steady changes of the interband Cooper scattering. On the other hand, the divergent length $l_d$ shows sensitive dependence in the regime $0<n<n_c$. This is expected because the (mean-field) transition temperature is roughly $T_c \sim e^{-l_d}$ by standard scaling argument. Compared with BCS theory, $l_d$ is inverse proportional to the density of states and thus proportional to the Fermi velocity (in quasi-one dimension). The almost-linear dependence of $l_d$ in Fig. 2 just reveals the underlying velocity evolution.

\begin{figure}
\centering
\includegraphics[width=7cm]{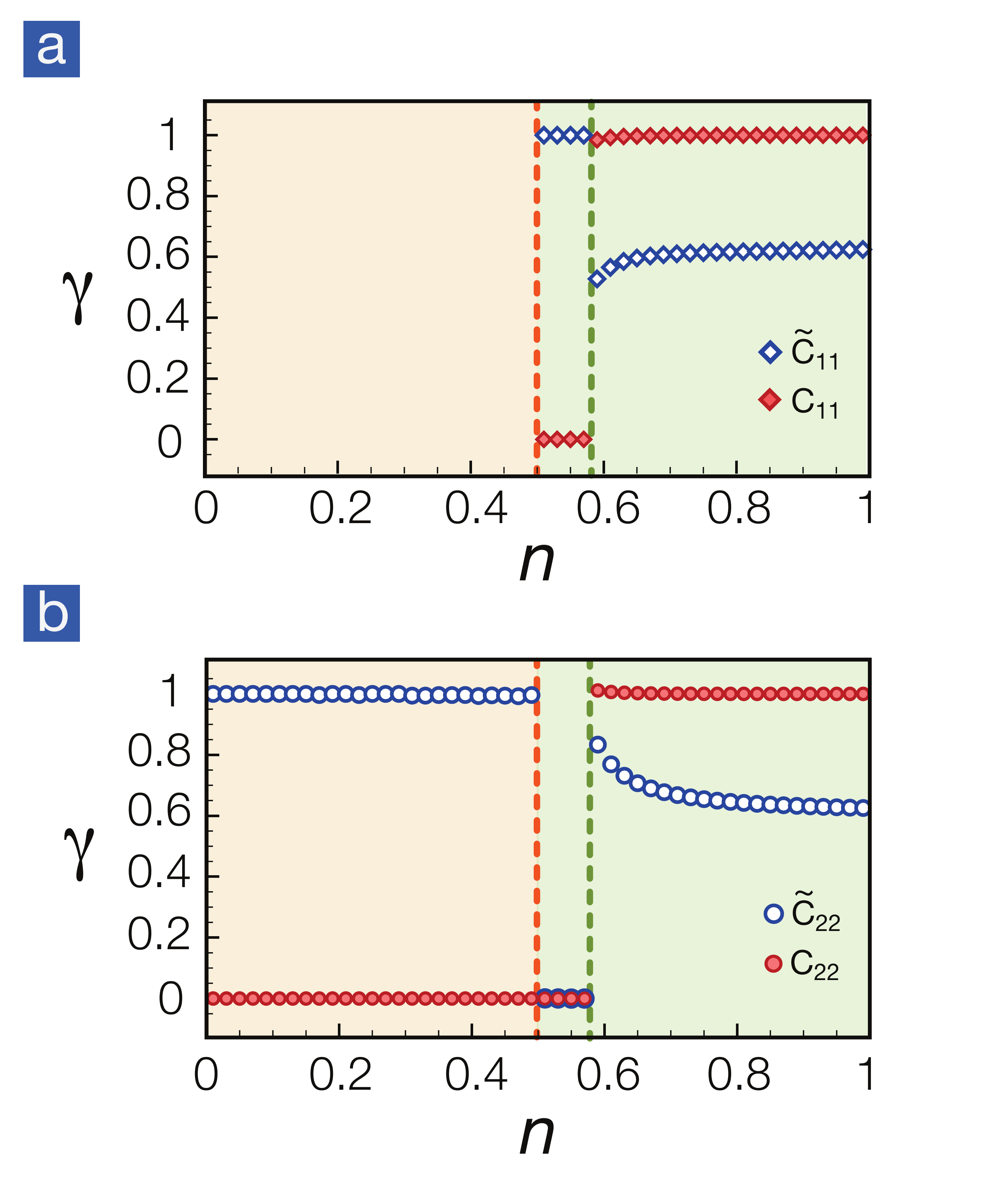}
\caption{RG exponents of intraband Cooper scattering in different superconducting phases (a) for the anti bonding band (b) for the bonding band. The pattern of the RG exponents provides a simple yet unique indicator to classify different ground states.
}
\end{figure}

Now we come to the RG exponents (shown in Fig. 3) in these superconducting phases. In the filling range $0.57 \lesssim n <1$, the RG exponents: $\gamma_{c_{11}}=\gamma_{c_{22}}=1$ are the largest among all and the pairing mechanism mainly arises from the electron-electron interactions. The RG exponent $\gamma_{\tilde{c}_{11}}$ is around $0.6$ near half filling $n=1$ and remains almost constant (but with slight decrease) upon filling reduction. On the other hand, $\gamma_{\tilde{c}_{22}}$ starts around $0.6$ near $n=1$ but increases to almost unity when approaching $n \approx 0.57$. The enhanced retarded Cooper scattering in the bonding band $\tilde{c}_{22}$ correlates with the larger spin gap from the electronic correlations. Although the phonon-mediated interactions are also relevant (positive RG exponents), the exponents $\gamma_{\tilde{c}_{11}}, \gamma_{\tilde{c}_{22}}$ in the retarded interactions are smaller than one and only play a secondary role in this regime. In short, one can say that the pairing mechanism in this regime is mainly attributed to electron-electron interactions. However, the observed physical quantities are ``dressed up" by retarded couplings and may depend on the electron-phonon interactions significantly.

With filling $0.5 < n \lesssim 0.57$, BCS instability occurs in the nearly empty band due to the Van Hove singularity. Note that this regime is almost twice larger than the 2-channel Luttinger liquid phase ($0.5<n\lesssim 0.54$) when the phonon-mediated interactions are ignored. As shown in Fig. 3, all RG exponents in this regime are zero except $\gamma_{\tilde{c}_{11}}=1$. The pattern of the RG exponents makes the analysis rather straightforward. Electrons form  pairs driven by the effective attraction in the (almost empty) antibonding band and other couplings remain small even at the cutoff length scale. The pairing mechanism is further supported by the linear dependence of the divergent length $l_d \propto (n-n_c)$ seen in Fig. 2. It is computed that the density of states in this regime is mainly ascribed to the antibonding band and the emergent Van Hove singularity takes the form $1/(n-n_c)$. Following BCS analysis, the critical temperature for the superconducting phase is $T_c \sim e^{-1/\tilde{g}} = e^{-{\rm const.} \times (n-n_c)}$. On the other hand, scaling argument gives $T_c \sim e^{-l_d}$. Comparing both approaches together, the linear dependence of the divergent length $l_d$ is explained.

Entering the one-band regime,$0<n<0.5$, the pattern of the exponents are also quite simple. As shown in Fig. 3, all RG exponents in this regime are zero except $\gamma_{\tilde{c}_{22}}=1$. The pairing instability now switches to the bonding band and the effective action of the ground state is well captured by BCS theory. The overall linear trend of the divergent length $l_d$ (shown in Fig. 2) follows the inverse of the density of the states as explained in the previous paragraph.

The pattern of RG exponents build up the hierarchy of relevant couplings and serves as an unambiguous indicator of the ground state. Furthermore, the pattern change is quite sharp when going through a quantum phase transition. The usage of these exponents are not limited to the simple two-leg system. We have applied the classification scheme to more general correlated systems in two dimensions and the primitive results suggest the RG exponents can also be extracted numerically. In fact, the proposed scaling ansatz has a solid root from mathematical aspect. For typical divergent flows studied here, a so-called $\psi$-series\cite{Goriely2000} is shown to exist with a single divergent length scale $l_d$. The RG exponent represents the leading order contribution of the infinite series expansion. Apparently, in most parameter regimes, the leading order term seems to grab the essential information in the RG flows. Due to the flexibility and generality of the $\psi$-series expansion, it is not yet clear whether the leading-order term can always be made dominant. At presence, the dominance of the leading order term is demonstrated in numerics without rigorous proof.

The dominant scaling ansatz also gives rise to non-trivial constraints between the RG exponents. For instance, plugging scaling ansatz for phonon-mediated intraband Cooper scattering $\tilde{c}_{ii}$ in Eq.~(\ref{rg-eqs}), it would imply either both $c_{ii}$ and $\tilde{c}_{ii}$ are irrelevant at the same time, or the exponents must satisfy the constraint, $\gamma_{\tilde{c}_{ii}}+1 
= \max \{ 2\gamma_{\tilde{c}_{ii}}, \gamma_{c_{ii}}+ \gamma_{\tilde{c}_{ii}}\}$. That is to say, the RG exponents for the intraband Cooper scattering are constrained,
\begin{align}
\gamma_{\tilde{c}_{ii}} = \gamma_{c_{ii}} =0,
\quad \mbox{or} \quad
\max \{ \gamma_{\tilde{c}_{ii}}, \gamma_{c_{ii}}\} = 1.
\end{align}
These constraints are apparent in the extracted RG exponents shown in Fig. 3. Other constraints on the RG exponents can be derived in similar fashion and agree with our numerical results at hand. Finally, we like to conclude that the scaling ansatz characterized by the divergent length $l_d$ and the RG exponents $\gamma_i$ also works for  phonon-mediated interactions. The classification scheme proposed here is simple but powerful without ambiguity.

We acknowledge supports from the National Science Council in Taiwan through grant NSC 100-2112-M-007 -017 -MY3. Financial supports and friendly environment provided by the National Center for Theoretical Sciences in Taiwan are also greatly appreciated.

\end{document}